# Retrieval-Augmented Sketch-Guided 3D Building Generation

*Generative Architectural Design for Japanese Detached Houses*

Zhengyang Wang[1], Nuttapong Rochanavibhata[2], Yuxiao Ren[3], Xusheng Du[4], Ye Zhang[3*] and Haoran Xie[5*]

[1,4,5]*Japan Advanced Institute of Science and Technology*
[2]*Mahidol University*
[3]*Tianjin University*
[5]*Waseda University*
xie@jaist.ac.jp
[*]*Corresponding authors*

**Abstract.** In the early design stage of Japanese detached houses, the lack of a unified design representation among clients, sales representatives, and designers leads to design drift and inefficient feedback. Usually, sketches handed off by sales representatives may lose details for quick drawing, which reduces the fidelity of subsequent 3D generation using generative AI models. The generated 3D model typically takes the form of a single unified mesh, preventing component-level editing. To solve these issues, we propose a multi-stage 3D generative design framework capable of producing architectural models from rough design sketches. The framework combines generative and retrieval-based methods to enable component-level editing and personalized customization. It adopts a multimodal representation for 3D model generation and applies component segmentation to localize architectural components such as windows and doors and uses retrieval to support targeted replacement of components. Experiments show that the work enables modular customization which is thought to be suitable for personalized architectural design. This work introduces a multi-stage sketch-to-3D framework for Japanese detached houses, provides facade and component datasets, and shows effectiveness through quantitative and expert evaluations.

**Keywords.** Architectural Design, Conditional Diffusion Models, Sketch-Based 3D Generation, Sketch-Based 3D Retrieval, Japanese Detached Houses

## 1. Introduction

In the design workflow of Japanese detached houses, the main challenge in the initial requirements-confirmation stage concerns the transmission and translation of intent among the client, sales representative, and designer rather than technical implementation. The coordination among them is usually constrained by limited visualization and communication skills. The client, as the source of requirements, conveys primarily intuitive preferences and lacks professional design knowledge and expressive skills such as spatial imagination and proportional awareness, which makes it difficult to describe details precisely, including openings and roof form. The sales representative, acting as an intermediary, translates these vague preferences into preliminary material intelligible to the designer. Without drawing training, requirements are captured as rough sketches with cluttered strokes, unclear outlines, and imbalanced proportions. However, in Japanese detached houses, the distinctive window-to-wall ratios and opening patterns require precise size and placement, which rough sketches do not convey reliably. As a result, architectural designers often need to model by hand from incomplete sketches, slowing the early design loop and causing frequent facade and model revisions.

Recent advances in generative AI have created opportunities to streamline sketch-to-3D generation (Li et al., 2024). However, current generative AI approaches for architectural design may remain misaligned with collaborative design scenarios. On the one hand, 3D generation methods typically rely on fine-grained image inputs (Koley et al., 2024). In practice, non-professional sketches often exhibit substantial loss of geometric information, making spatial semantics difficult to infer and yielding 3D results that deviate from design intent. On the other hand, the generated 3D output typically takes the form of a single unified mesh without component-level segmentation, preventing independent identification and editing of key components such as doors and windows. When local replacement is requested, the designer still need to reconstruct meshes manually.

To address these challenges, this paper proposes a multi-stage 3D generative design framework tailored to the early design stage for Japanese detached houses, enabling automated conversion from rough 2D hand-drawn sketches to 3D models with component-level editability, including doors and windows. This paper is positioned as a design-support tool for professional designers to streamline model construction in early-stage communication, not a substitute for downstream design decisions such as structural optimization and material selection. Specifically, the framework proceeds through the following three stages. In the sketch-to-image enhancement stage, ControlNet (Zhang et al., 2023) uses the rough sketch as a conditioning input to generate an information-rich intermediate facade image that mitigates blurred geometry and missing textures and supports subsequent 3D reconstruction. In the 3D model generation stage, the enhanced image is fed into the state-of-the-art image-to-3D reconstruction model (Xiang et al., 2025), which transforms 2D spatial features into a 3D structured mesh. In the sketch-guided retrieval and replacement stage, Grounded-SAM (Ren et al., 2024) generates 2D masks of target components on 2D projections of the 3D model. The masks are mapped to the 3D mesh under the camera projection to segment components precisely. Sketch-based model retrieval (Eitz et al., 2012) extracts geometric features from provided reference sketches of target components and retrieves, for each target component, the best-matching 3D component from an external architectural-component dataset. Finally, spatial alignment and mesh fusion integrate the selected component into the corresponding segmented region, ensuring structural continuity while enabling efficient component replacement and rapid iterative design.

The main contributions of this study are summarized as follows:

- We propose a multi-stage 3D generative design framework that produces component-editable 3D architectural models from rough 2D sketches, with a focus on Japanese detached houses.

- We establish two datasets for Japanese detached houses: a facade dataset to enhance style alignment, and an architectural component dataset for retrieval-based component replacement.

- We provide quantitative analyses and expert evaluations demonstrating the effectiveness of the proposed framework.

## 2. Literature Review

This section surveys prior work relevant to this study. We first discuss sketch-to-image generation and retrieval-augmented approaches for architecture image generation, and then examine recent studies in 3D model generation and reconstruction.

### 2.1. SKETCH-TO-IMAGE GENERATION

Recent research on conditioned generation has become an important direction for automating architectural design (Li et al., 2024). Within this line of work, conditional diffusion models such as ControlNet (Zhang et al., 2023) incorporate conditioning information during generation (e.g., sketches and depth maps) to impose spatial constraints. This conditioning helps maintain structural features while improving perceptual fidelity. However, robustness in many methods depends on large, high-quality training datasets, which limits practicality when clean data are scarce. This study uses Low-Rank Adaptation (LoRA) (Hu et al., 2022) fine-tuning to enable small-sample training, reducing dependence on extensive datasets while producing fine-quality Japanese detached house images suitable for downstream structural inference.

Retrieval-augmented generation (RAG) (Lewis et al., 2020) introduces external datasets or exemplar libraries to provide additional knowledge or conditioning signals for generative models. In architectural design, prior studies (Wang et al., 2025) retrieve local sketch regions from external datasets to select objects that best match target forms or semantics, which enriches content and promotes style consistency. However, in 3D architectural generation, establishing spatial correspondence between retrieved objects and specific local components and performing precise replacement remains challenging.

### 2.2. 3D MODEL GENERATION

Neural 3D reconstruction began with radiance-field formulations that learn view-consistent volumetric representations from multi-view images, as represented by NeRF (Mildenhall et al., 2021) and its successors, which improved anti-aliasing, surface recovery, and robustness under sparse views. Efficiency was later advanced by fast radiance representations such as 3D Gaussian Splatting (Kerbl et al., 2023), which enable real-time view synthesis while still relying on multi-view images. With diffusion models and large-scale priors, high-quality single-view 3D reconstruction has become feasible (Tochilkin et al., 2024), and some methods further enable editing directly in a structured 3D latent space (Xiang et al., 2025). However, the latent space provides neither per-part semantics nor precise placement, leaving edits unattached to specific architectural components. In addition, the representation is not directly visualizable and decoding typically produces a single unified mesh. These issues make consistent, fine-grained iteration difficult.

## 3. Method

This work proposes a three-stage generative framework to achieve component-level editability in sketch-based 3D architectural generation, as shown in Figure 1. First, the sketch is used as the conditioning input to generate a detailed, style-consistent 2D intermediate facade image. Next, a state-of-the-art diffusion-model-based image-to-3D model is adopted to generate a textured 3D architectural mesh from the intermediate facade image. Finally, component segmentation outputs masks for key components, which are then used to localize their 3D regions for editing. Given a reference component sketch, the framework retrieves the most relevant 3D component from an external dataset. Through spatial alignment and mesh fusion, it is inserted into the localized region, supporting modular customization and faster iteration.

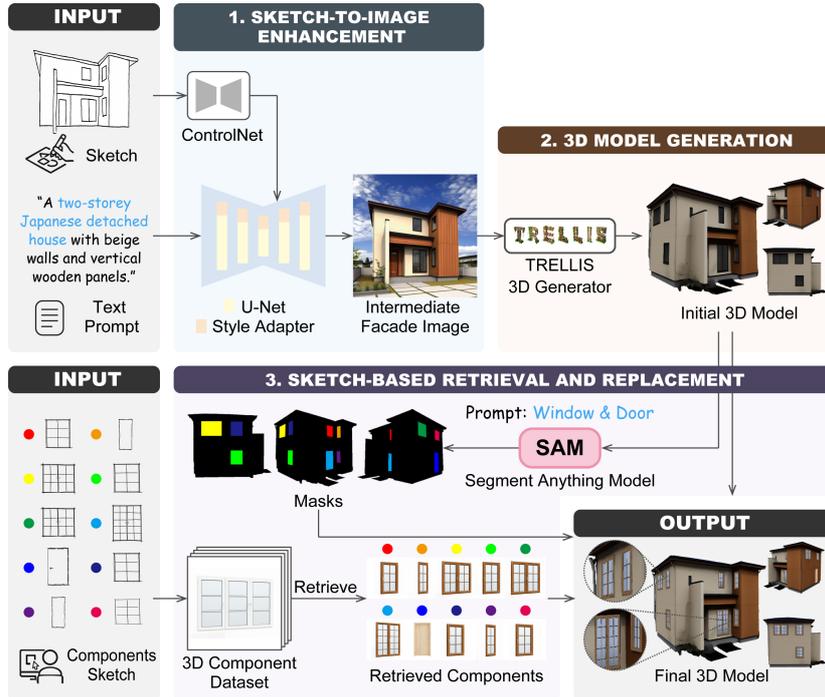

Figure 1. The overview of framework. Three-stage framework for component-level editing from sketches: sketch-to-image enhancement, 3D building model generation, and sketch-based retrieval and replacement.

# RETRIEVAL-AUGMENTED SKETCH-GUIDED 3D BUILDING GENERATION

### 3.1. DATASETS FOR JAPANESE DETACHED HOUSE MODELING

To achieve stable architectural style alignment for Japanese detached houses, a facade image dataset was constructed. Samples were drawn primarily from public real-estate listing pages on Ichijo Komuten (Ichijo Co., Ltd, n.d.) and SUUMO (SUUMO, n.d.), as shown in Figure 2. After collection, images underwent denoising and cleaning. Night scenes, strong specular reflections, severe occlusions, and multi-building compositions were removed. To provide usable text conditions, text prompts were generated with the vision-language model BLIP-2 (Li et al., 2023) and then manually verified by professionals with an architectural background. Given LoRA's robustness with small sample sizes and low training cost, a compact and curated dataset was adopted. The final dataset contains 100 high-quality facade images.

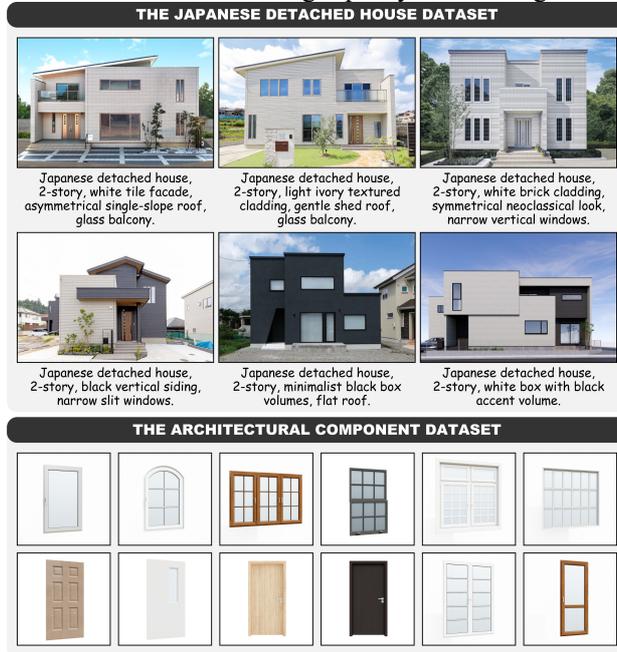

Figure 2. Overview of the datasets, including the Japanese detached house dataset and the architectural component dataset.

Auto-generated 3D architectural models often exhibit inconsistencies or uneven quality in details such as doors and windows. To support component-level replacement, a 3D component dataset was constructed for the facades of Japanese detached houses, comprising 400 3D component models. Style and specification choices were guided by the product lines of market-leading Japanese fenestration manufacturers, covering representative types across sizes, materials, and styles, as illustrated in Figure 3. Models are stored in glb format. Each glb file bundles geometry, materials, and textures, enabling efficient loading and precise matching throughout the retrieval-and-replacement framework.

### 3.2. SKETCH-TO-IMAGE ENHANCEMENT

The rough sketches may be not suitable to guide 3D building generation directly because of imbalanced proportions, discontinuous strokes, and missing details. To address this issue, a facade image is adopted as intermediate between the sketch and the 3D building reconstruction. We adopt LoRA (Hu et al., 2022) to fine-tune the ControlNet (Zhang et al., 2023) backbone U-Net on our constructed facade dataset of Japanese detached houses, achieving style alignment. ControlNet then uses the geometry of the input sketch $S$ as the structural basis. The style adapter together with the sketch-conditioned ControlNet enables the sampling process to maintain architectural details and style aligned with the text prompt. The predicted noise at the current timestep is given as:

$$y_c = \mathcal{F}(x; \Theta + \Delta\Theta) + \mathcal{Z}(S; \Phi),$$

where $\mathcal{F}$ is the ControlNet backbone U-Net. $\Theta$ are its frozen base parameters. $\Delta\Theta$ is the LoRA increment applied to the backbone. $x$ is the latent representation in the diffusion process. $\mathcal{Z}(S; \Phi)$ is the ControlNet conditioning branch, $S$ is the input sketch and $\Phi$ denotes its parameters.

### 3.3. 3D BUILDING MODEL GENERATION

At this stage, the intermediate facade image is fed into state-of-the-art 3D generation method, TRELLIS (Xiang et al., 2025), a diffusion-based single-view image conditioned 3D generator, which outputs a 3D building mesh. The image is encoded as condition inputs to extract conditioning features that guide sampling in a structured 3D latent space. Denoising operates in the 3D latent space, and a textured 3D building mesh is recovered by the decoder. The mesh preserves architectural massing and the geometry of components.

The 3D generator employs an explicit sparse 3D representation that associates latent features with their voxel coordinates. This mechanism supports localized spatial editing in the latent space. However, edits are applied by region-wise resampling rather than replacement, and no constraints on scale or location are enforced. In addition, the output is a single unified mesh without part boundaries, instance labels, or stable correspondences. These factors hinder reliable component replacement and consistent iteration. Component-level editability is introduced in the subsequent segmentation and replacement stage.

### 3.4. SKETCH-BASED RETRIEVAL AND REPLACEMENT

At this stage, Grounded-SAM (Ren et al., 2024) segments architectural components such as doors and windows from text prompts, producing 2D masks that localize each component. Meanwhile, a depth map is rendered from the same viewpoint via OpenGL depth-buffer readback, which retrieves per-pixel depth values from the active framebuffer. The mask is matched to the depth map using the camera extrinsic parameters. Within each masked region, pixels with depth are retained as the foreground set. Using depth values and the camera extrinsic parameters, the foreground pixels are projected to world coordinates to obtain the component's 3D point cloud. From the 3D point cloud, an oriented bounding box (OBB) (Seidel, 1981) is fitted. Its freely oriented axes follow the cloud's principal directions and provide the centre, principal axes, side lengths, and eight vertex coordinates. In parallel, the reference component sketch is fed into the sketch-based image retrieval approach (Eitz et al., 2012) to retrieve the external component dataset and return the most relevant corresponding 3D component mesh. The candidate component's initial position and orientation are determined by the OBB centre and principal axes, and its scale is derived from the box side lengths under either uniform or per-axis scaling, thereby achieving alignment with the target region and replacing the mesh faces in that region with the candidate component. Each segmented component can be replaced, enabling modular customization while maintaining visual and geometric continuity. Figure 3 shows several component-editable 3D models generated by the framework.

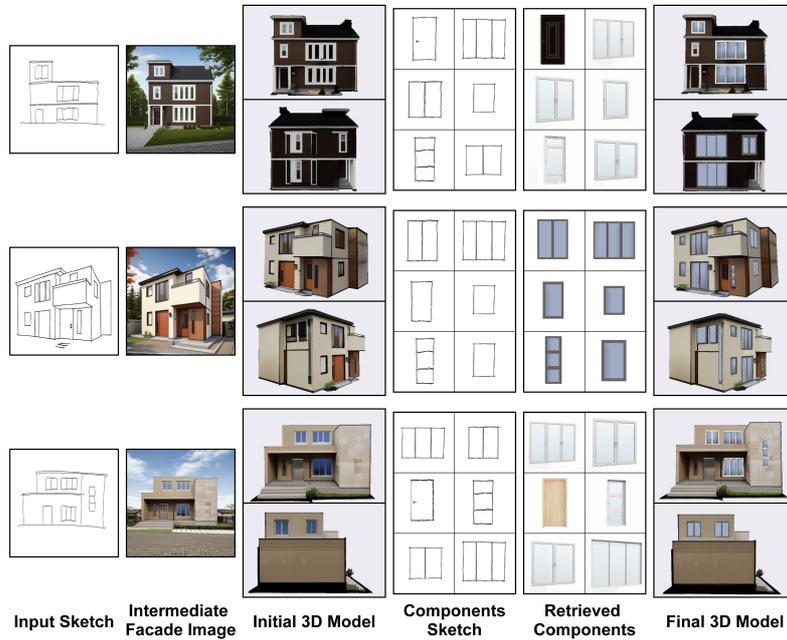

Figure 3. The results show the conversion from sketches to component-editable 3D models.

## 4. Evaluation

Quantitative evaluation used the same set of non-professional sketches as input, comparing three end-to-end pipelines: (1) basic 3D generation (w/o style adaptation), (2) style-enhanced 3D generation (w/o component replacement), and (3) ours (Full Pipeline). Note that a direct sketch-to-3D pipeline was excluded, as its grayscale outputs lack visual equivalence to image-conditioned results. Figure 4 illustrates qualitative examples comparing the three pipelines. To ensure comparability, the same text prompt and a fixed random seed were applied to all pipelines for each sketch. Two metrics based on CLIP (Radford et al., 2021) were adopted: CLIP-T for text–image similarity and CLIP-I for image–image similarity. Specifically, for each sketch, each pipeline generated a 3D architectural model, and a fixed, evenly spaced camera set rendered multi-view images from each model. CLIP-T measured text–image similarity by encoding the text prompt and, per view, encoding the rendered images, followed by cosine similarity. CLIP-I measured the agreement between the front-view rendering image and the intermediate facade image. To improve fairness and stability, affine alignment via the OpenCV-based image alignment method was used to mitigate differences due to scale and cropping, and cosine similarity between image embeddings was computed after alignment. As shown in Table 1, pipeline (2) style-enhanced 3D generation improves CLIP-I score over pipeline (1) basic 3D generation, suggesting that style adaptation helps transfer facade cues from the intermediate image to the reconstructed model. The full pipeline (3) shows the highest CLIP-T score and keeps CLIP-I score close to pipeline (2). The results indicate style adaptation improves the preservation of visual cues, and component replacement improves the clarity and consistency of components without disrupting facade structure.

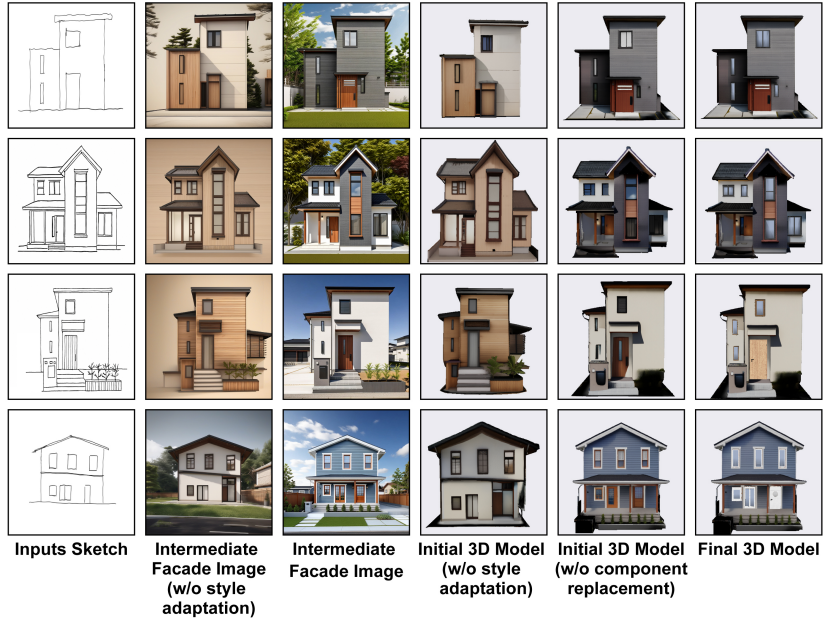

Figure 4. Qualitative comparison of three pipelines: (1) Basic 3D Generation (no style adaptation), (2) Style-Enhanced 3D Generation (style adapter only), and (3) Ours (Full Pipeline) (style adapter with component replacement).

Table 1. Quantitative analyses (CLIP-T/CLIP-I) and human evaluation

| Method | CLIP-T ↑ | CLIP-I ↑ | Visual Quality | | Sketch Consistency | | Text Alignment | |
|---|---|---|---|---|---|---|---|---|
| | | | Mean | Std | Mean | Std | Mean | Std |
| (1) Basic 3D Generation | 0.280 | 0.883 | 3.339 | 0.959 | 3.366 | 0.999 | 3.419 | 0.987 |
| (2) Style-Enhanced 3D Generation | 0.279 | **0.908** | 3.820 | 0.781 | 3.589 | 0.796 | 3.991 | 0.773 |
| (3) Ours (Full Pipeline) | **0.285** | 0.906 | 4.169 | 0.801 | 4.152 | 0.805 | 4.330 | 0.719 |

To further assess the generative capability of the proposed framework, a human evaluation was conducted with 56 participants possessing backgrounds in architecture and design. Using a five-point Likert scale ranging from 1 (very poor) to 5 (excellent), the evaluators rated the outputs of three different generation pipelines corresponding to each input sketch across three aspects: overall visual quality, consistency with the input sketch, and alignment with the input text prompt. In the human evaluation, the full pipeline obtained a mean visual-quality score of 4.169 (SD = 0.801), where higher values indicate better perceived realism, clarity, and completeness. For sketch consistency, the mean was 4.152 (SD = 0.805), where higher values indicate closer correspondence to the input sketch in layout, proportions, and component placement. For text alignment, the mean was 4.330 (SD = 0.719), where higher values indicate stronger agreement with the text prompt in style and component attributes.

## 5. Conclusion

This work addresses the challenges of intent drift and inefficient design iteration during the early requirement-confirmation stage of Japanese detached housing by introducing a multi-stage generative design framework that enables component-level editability. The framework is positioned as a design-support tool for professional designers in early-stage communication rather than a substitute for downstream design decisions such as structural optimization and material selection. To implement this framework, two datasets are established: one comprising facade images for style adaptation and another consisting of 3D components to facilitate replacement. Experimental evaluations demonstrate that, with lightweight training focused on style adaptation, the proposed framework effectively transforms non-professional sketches into text-aligned, component-editable 3D architectural models, thereby enhancing communication reliability and enabling more efficient.

This study also has several limitations. This work did not repair gaps at replacement boundaries or discontinuities in normals and texture coordinates (UVs), which may lead to shading artifacts, visible seams, and unnatural material transitions. In addition, the current framework did not support deletion of existing 3D components, which limits workflows that require removing elements before replacement or reconfiguration. As future work, we would like to introduce differentiable re-topology and automatic UV re-remapping to achieve seamless mesh fusion with consistent normals and UVs. Furthermore, the framework will be integrated with Building Information Modeling (BIM) workflows to enrich the generated geometry with construction and component metadata, thereby enabling a seamless transition from conceptual design to downstream engineering implementation.

### Acknowledgements

Zhengyang Wang was supported by the China Scholarship Council (CSC). This work was supported by JST BOOST Program Japan Grant Number JPMJBY24D6, the National Natural Science Foundation of China Grant Number 52508023, and JST SPRING, Japan Grant Number JPMJSP210.

### Attribution

We used ChatGPT (OpenAI, 2024) to improve the language clarity and grammar, and Grammarly (Grammarly Inc., 2024) to correct spelling errors and typos.